\documentclass[aps,prl,reprint,amsmath,amssymb]{revtex4-2}
\usepackage{graphicx}
\usepackage{dcolumn}
\usepackage{bm}
\usepackage{changes}

\begin{document}

\preprint{APS/123-QED}

\title{Randomization of a Laser Wavefront by the Turbulent Gas-Puff \\ Z-Pinch Plasma Column}

\author{A. Rososhek}
    \email{ar877@cornell.edu}
\author{E. S. Lavine}
\author{B. R. Kusse}
\author{W. M. Potter}
\author{D. A. Hammer}
\affiliation{Laboratory of Plasma Studies, Cornell University Ithaca, New York 14853, USA
}%

\begin{abstract}
In this paper, we present the first direct experimental evidence supported by numerical modeling of a turbulent plasma column formed during a gas-puff z-pinch implosion generated by the COBRA current. Utilizing an imaging refractometer, we showed a significant decrease in spatial autocorrelation of the laser field and the appearance of a laser speckle pattern shortly before stagnation. The intensity distribution of the speckles measured during different shot campaigns while employing long and short COBRA pulses follows the speckle statistics satisfactorily. The imaging refractometer signal is proportional to the integral over the electron density gradients; hence, the measured phase randomization of the individual plane waves comprising the laser field implies a random density distribution. To validate this, the Beam Propagation Method code simulates the laser beam propagation through different artificial density distributions with various average fluctuation scales and generates synthetic imaging refractometer data. The results reproduce similar trends in the experimental data, such as the increasing vertical width for the decreasing average spatial scale of the fluctuations and the decreasing spatial correlation length of the laser field. Therefore, during the gas-puff z-pinch implosion process, it is likely that the plasma flow is almost always turbulent with the average spatial scale of the turbulent density fluctuations decreasing towards stagnation.
\end{abstract}
\maketitle

\par Turbulence in high-energy-density (HED) plasma flows has been continuously discussed at different levels, from the theoretical perspective to the experimental evidence \cite{PaulDrake_2008,White2019,Collins_20}. The Kelvin-Helmholtz, Rayleigh-Taylor (including magneto-Rayleigh-Taylor (MRT)), and Richtmyer-Meshkov instabilities are all known to create conditions for the flow to become turbulent\cite{OHarricane_turb, Nagelturb, Weber_RMmix}. A detailed understanding of the relationship between these instabilities and turbulent mixing in the HED environment is crucial from theoretical and applied perspectives.
\par Recently, Kroupp et al. reported that while reviewing their older spectroscopy data, they achieved a better fit and a generally more physically sound picture by assuming the plasma density distribution to be supersonically turbulent \cite{Kroupp_turbo,Kroupp_measure}. They found that moving from a uniform to a turbulent density distribution accounts for half the average inferred density and better agrees with the experiment while assessing the emitting body size (the plasma column radius). The assumption of turbulence was primarily based on the estimated Reynolds number of approximately $10^5$ and a Mach number greater than 1. While these estimates are indicative of turbulence, the Reynolds number calculation relies on undisclosed assumptions about plasma viscosity, which depends on plasma transport properties and charge state, rendering it somewhat ambiguous. Further research involving the electron plasma wave feature in the time-resolved and the ion acoustic wave feature in the time-integrated Thomson scattering spectrum gave a more detailed picture \cite{Rocco_epw, Sander_iaw}. It was shown that to fit the measured spectra an additional non-thermal Gaussian broadening term should be introduced. This broadening peaks at the front of the plasma sheath, and its scaling behavior is inconsistent with laminar velocity gradients for the chosen size of the collection volume, which may exclude the effects of smaller-scale velocity gradients, if present. Furthermore, upstream ion penetration length studies indicate collisionless interaction within the plasma sheath. The data from Ref.~\cite{Rocco_epw} suggest a time frame of at least $40~\text{ns}$ before the stagnation column formation for the onset of turbulence; however, this conclusion relies on a low-resolution dataset of the electron plasma wave spectral feature, as reflected in the $\chi^2$ statistics of the fits. Thus, given all the indications, there is no direct experimental or numerical evidence to support the hypothesis of turbulent gas-puff implosions. The question of when and how the imploding plasma flow becomes turbulent remains unresolved. In this paper, we would like to present the first direct experimental evidence of turbulence supported by numerical simulations. The primary diagnostic tool used for this research is an imaging refractometer technique complemented by the Beam Propagation Method (BPM) simulation \cite{Jhare_imref, okamoto2006}.
\begin{figure*}[htbp]
    \centering
    \begin{minipage}[t]{0.32\textwidth}
        \centering
        \includegraphics[height=1in]{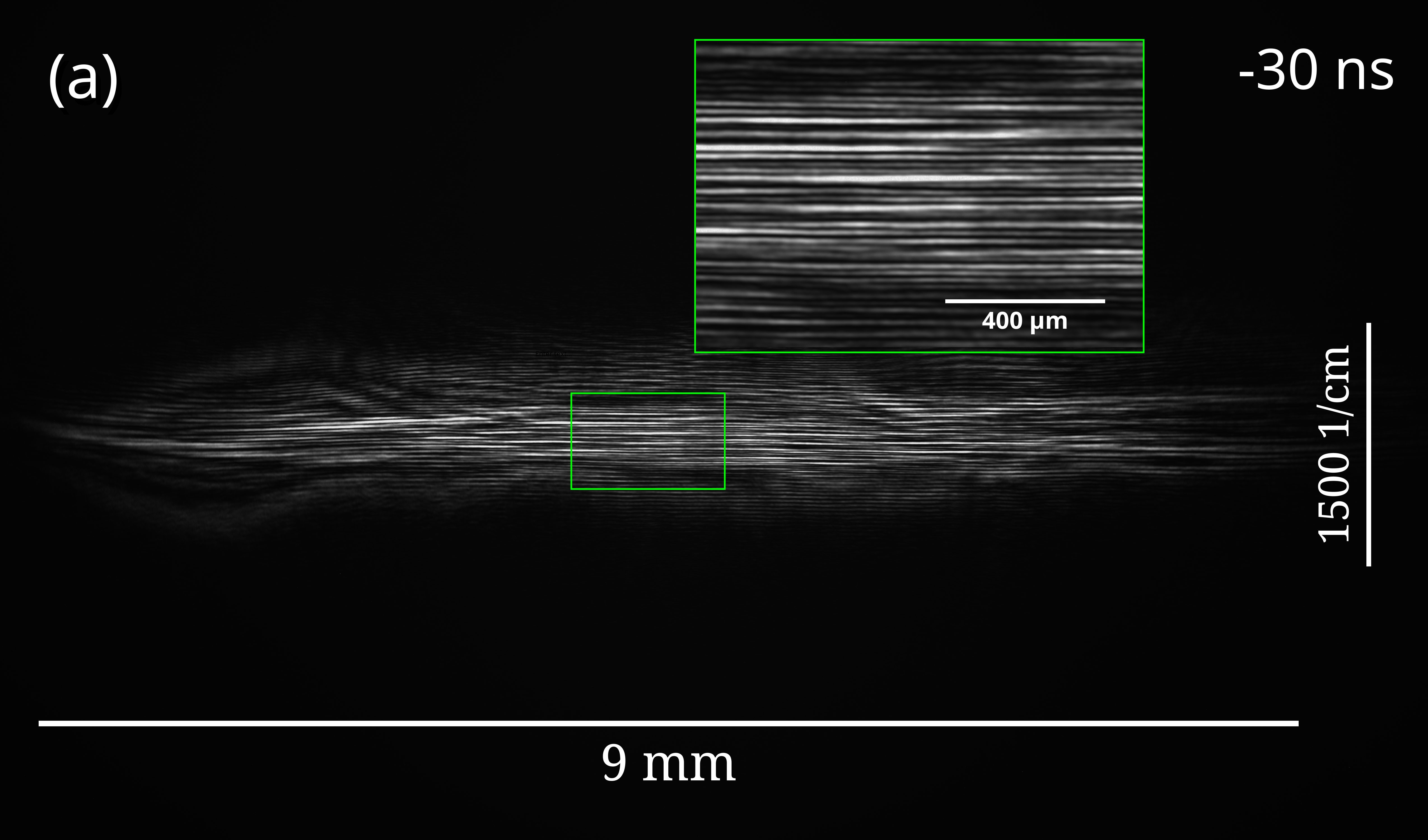}
    \end{minipage}
    \hfill
    \begin{minipage}[t]{0.32\textwidth}
        \centering
        \includegraphics[height=1in]{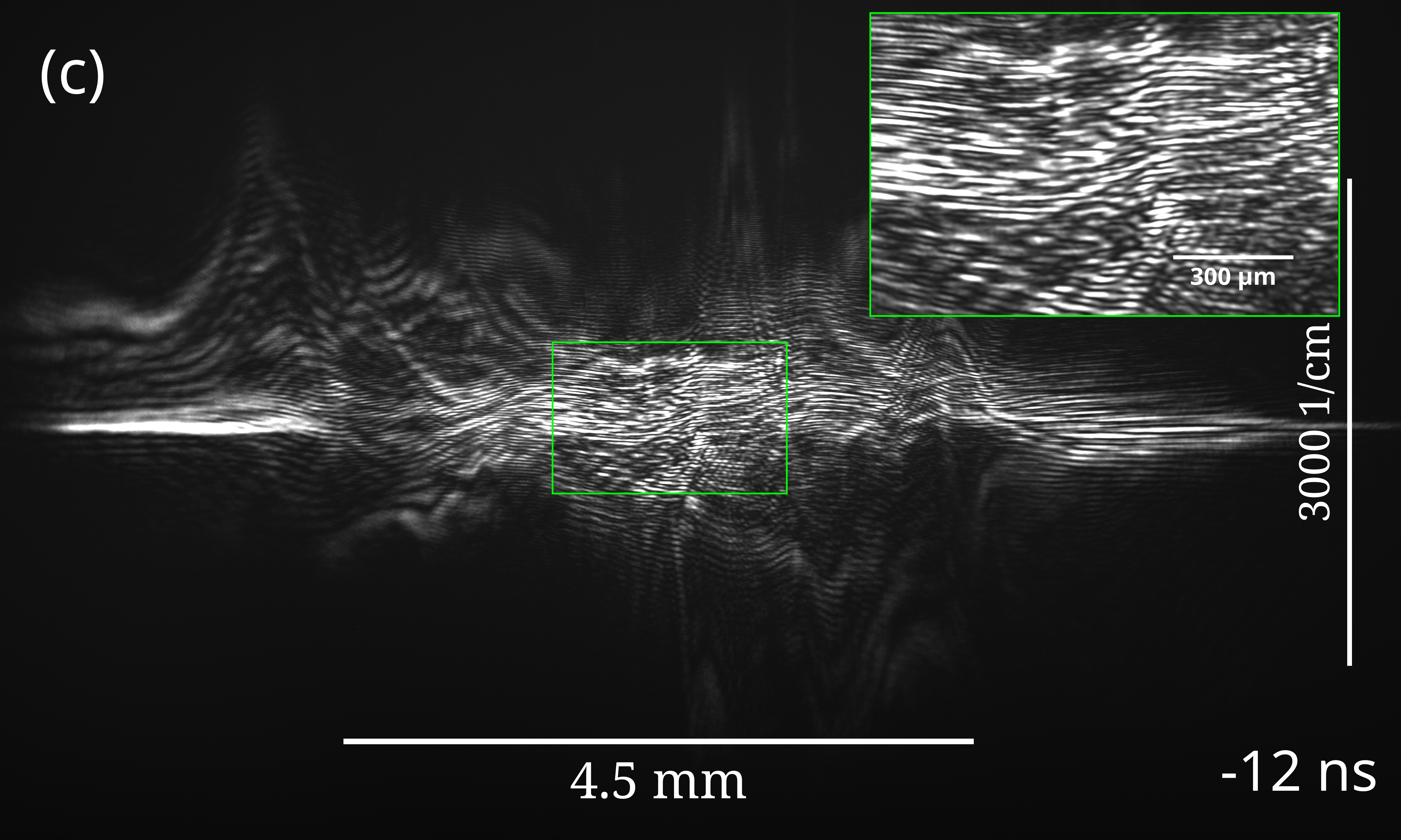}
    \end{minipage}
    \hfill
    \begin{minipage}[t]{0.32\textwidth}
        \centering
        \includegraphics[height=1in]{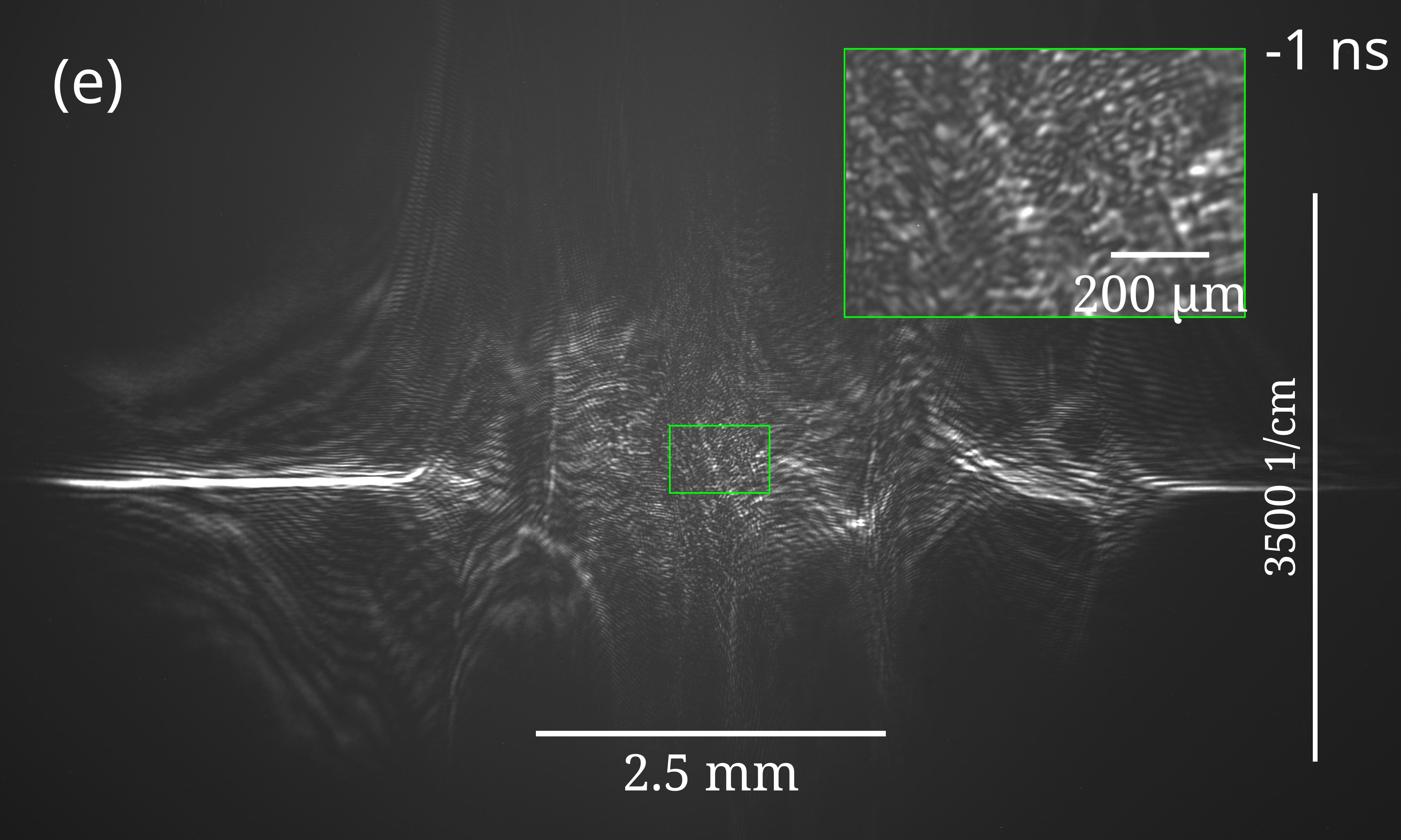}
    \end{minipage}
    \\
    \vspace{0.4cm}
    \begin{minipage}[t]{0.32\textwidth}
        \centering
        \includegraphics[height=1in]{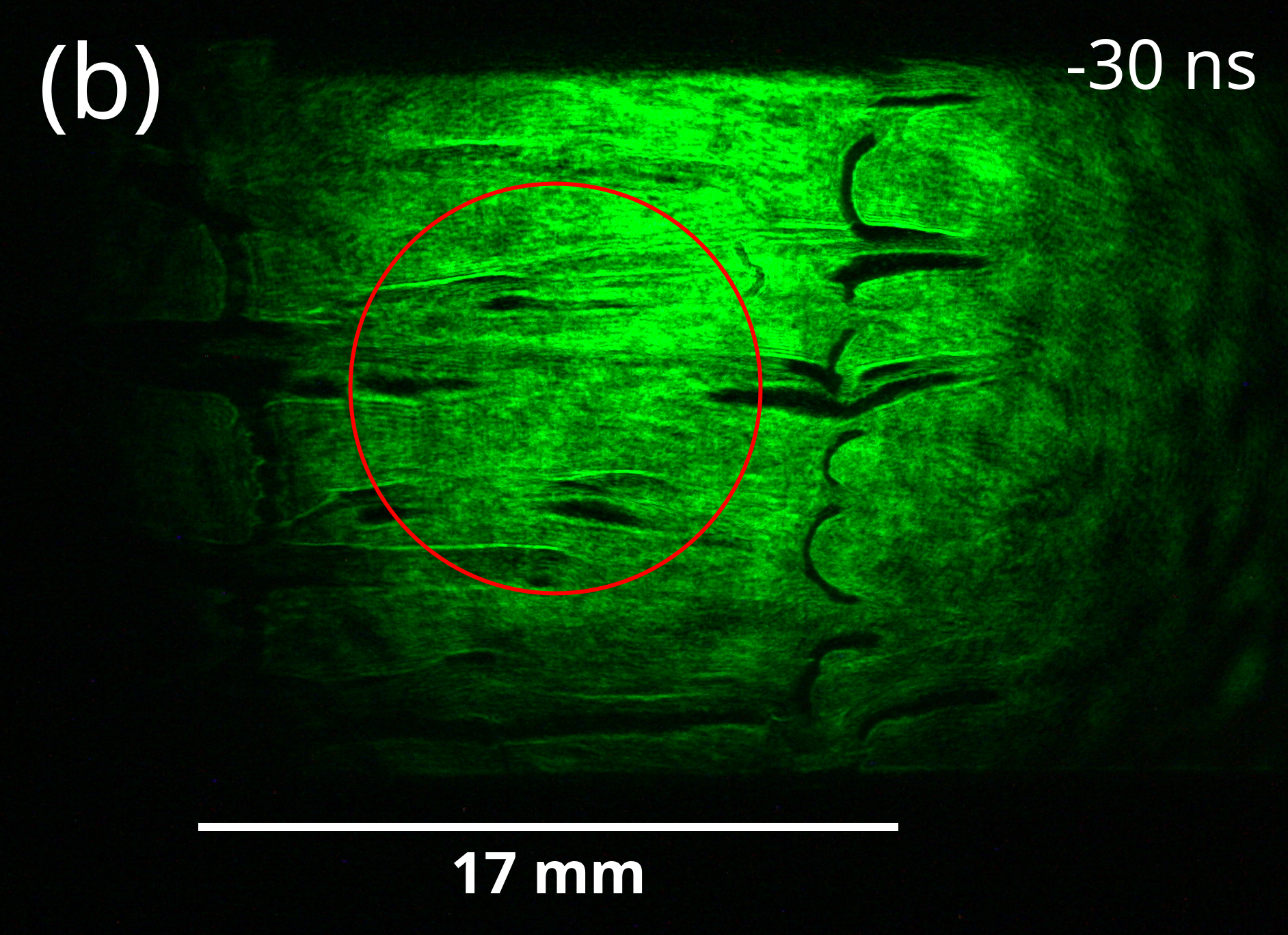}
    \end{minipage}
    \hfill
    \begin{minipage}[t]{0.32\textwidth}
        \centering
        \includegraphics[height=1in]{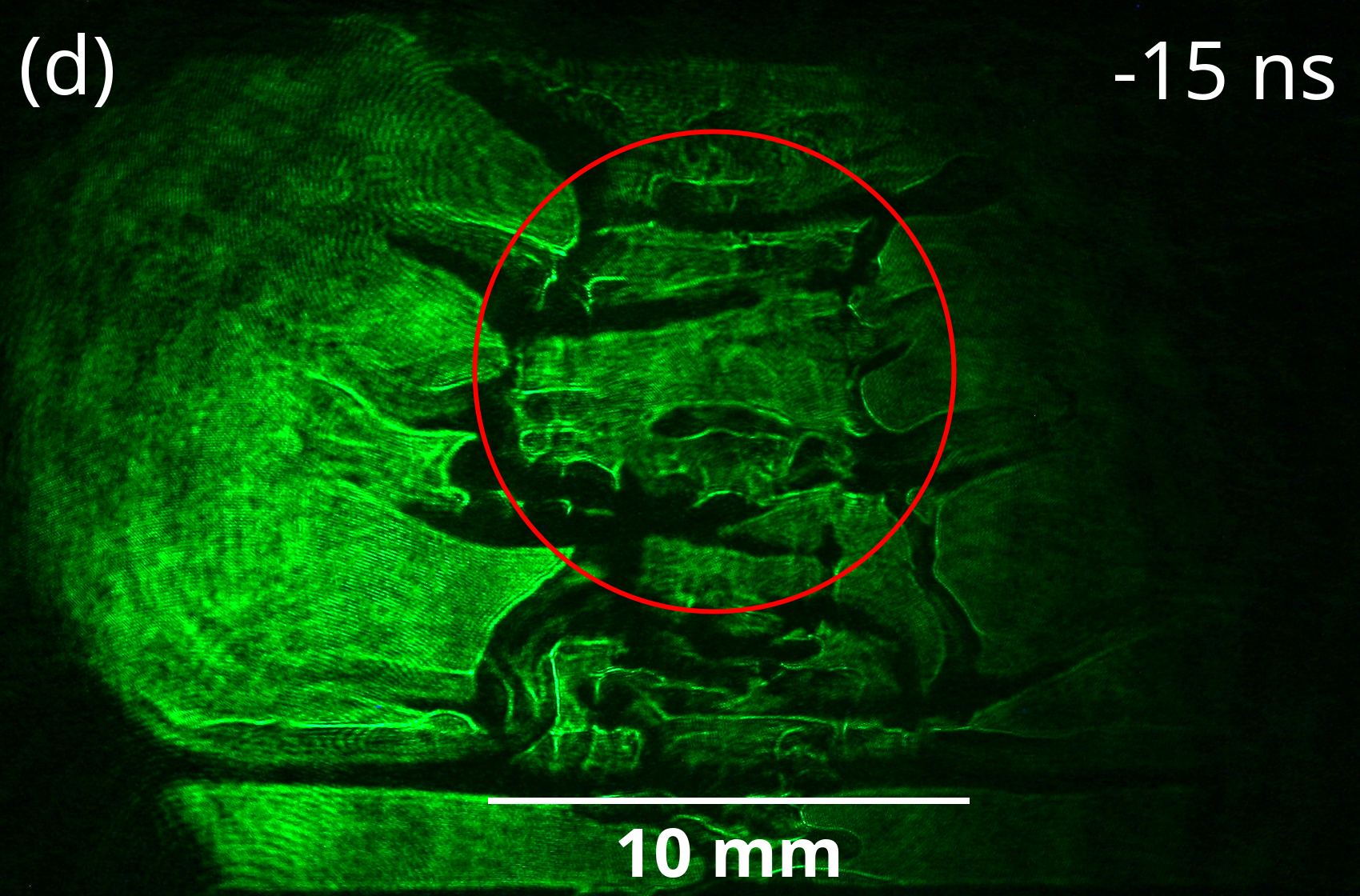}
    \end{minipage}
    \hfill
    \begin{minipage}[t]{0.32\textwidth}
        \centering
        \includegraphics[height=1in]{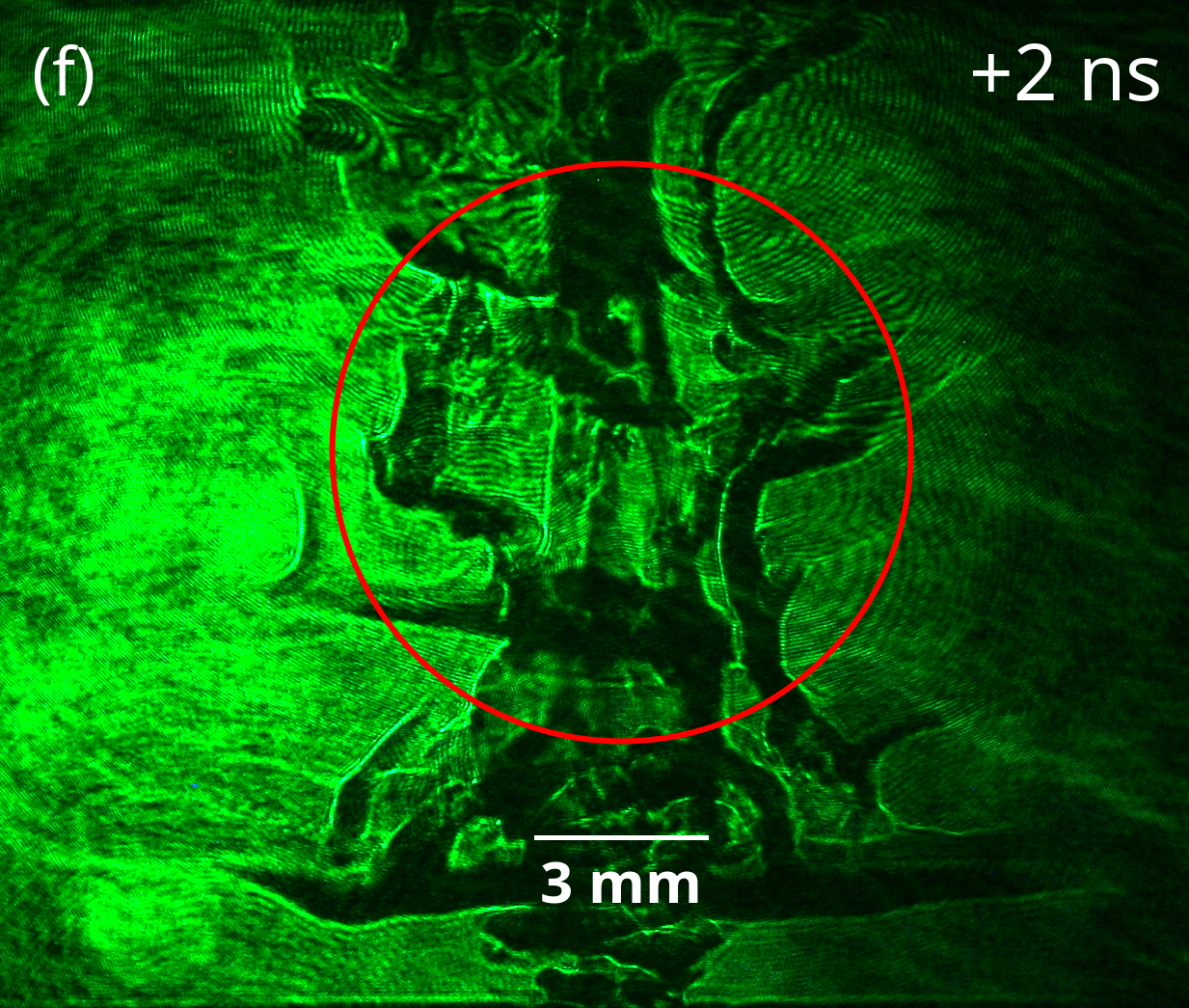}
    \end{minipage}
    \caption{The experimentally measured imaging refractometer and shadowgraphy images, including zoom-in regions marked in green, where (a, c, e) represent raw imaging refractometer data from shots \#7007, \#7019, \#7009, respectively, and (b, d, f) - typical shadowgraphy images for the observed plasma column diameters with a red circle to show approximately where the imaging refractometer beam was directed. The timings are relative to the x-ray emission maximum measured by a PCD.}
\label{fig:imref_shdw}
\end{figure*}

\par In what follows, we will describe the experimental platform, the diagnostics used, and the BPM simulation. Afterward, the experimentally measured data that supports turbulent hypothesis will be presented and analyzed, followed by discussion of the BPM simulations.

\par The experimental platform to drive the implosions is the COBRA generator \cite{Cobra}, with a gas-puff load delivered via an independently pressurized triple nozzle valve designed in collaboration with the Weizmann Institute. The COBRA pulsed power generator features energy storage provided by two Marx generators, each containing $16\times1.35\mu F$ capacitors charged to 70 kV, yielding approximately $105\!kJ$ of stored energy. It is a low-impedance ($0.5\Omega$) machine designed to drive a $1\!MA$ current through loads with an inductance on the order of $10nH$, with a variable rise time of approximately $100\!ns$ (short pulse) and $230\!ns$ (long pulse). In the current experiments, a Neon machine over-massed load ($\approx\!44~\text{$\mu$g/cm}$, $\approx\!25~\text{cm}$ long) was utilized with both long and short COBRA pulses. A photoconducting diamond (PCD) detector is used along with a $6~\text{$\mu$m}$ Mylar filter to monitor X-ray emission from the stagnation column. Two $\approx\!140~\text{mJ}$, $150~\text{ps}$, frequency-doubled Nd:YAG laser pulses at 532 nm are employed for shadowgraphy and imaging refractometry \cite{Jhare_imref}. The imaging refractometer is a laser wavefront probing diagnostic that allows deflection angles---proportional to the density gradient in a thin plasma limit---to be measured on the $y-axis$ while spatial information, in this case chosen to be the pinch radius, is along the $x-axis$. The imaging refractometer utilizes a spatially filtered, single TEM\textsubscript{00} mode laser profile of $\approx\!\!9~\text{mm}$ size and centered at the $r=0$ position of the plasma column while viewing the middle of the cathode-anode plane. In this setup, the laser beam propagates across the pinch column diameter in the $r-z$ plane of the cylindrical coordinate system. The shadowgraphy laser is about $50~\text{mm}$ in size, is not spatially filtered and is used for viewing the whole cathode-anode gap region of 45$^\circ$ relative to the imaging refractometer, and is synchronized to fire almost simultaneously with it. We used a QHY268M-Pro camera with a very low readout noise 16-bit depth CMOS sensor and an interference filter for the imaging refractometer measurements. We utilized a Canon DSLR camera with a polarizer-interference filter pair for the shadowgraphy to reduce plasma self-emission. In the presented research, the laser beams are injected into the experimental chamber and propagate through the central jet at $r=0$, similar to the scheme described in Ref.~\cite{Sander_iaw}. This work shows typical current waveforms, Thomson scattering data, and different extreme ultraviolet images for different load types. The experimental campaign aimed to use the diagnostics above to observe different phases of the imploding flow.
\begin{figure*}[htbp]
    \centering
    \begin{minipage}[t]{0.32\textwidth}
        \centering
        \includegraphics[,height=1.2in]{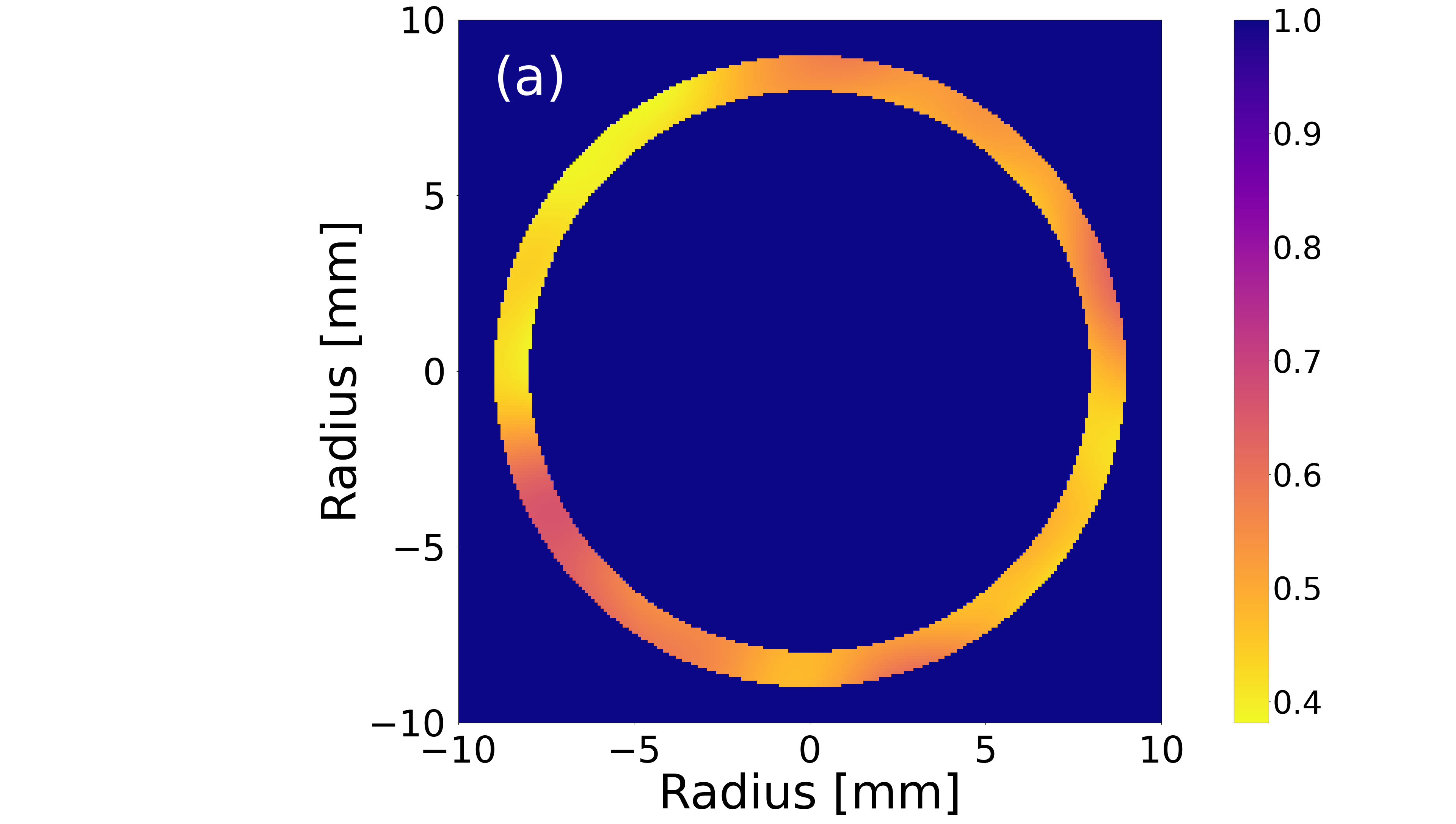}
    \end{minipage}
    \hfill
    \begin{minipage}[t]{0.32\textwidth}
        \centering
        \includegraphics[height=1.2in]{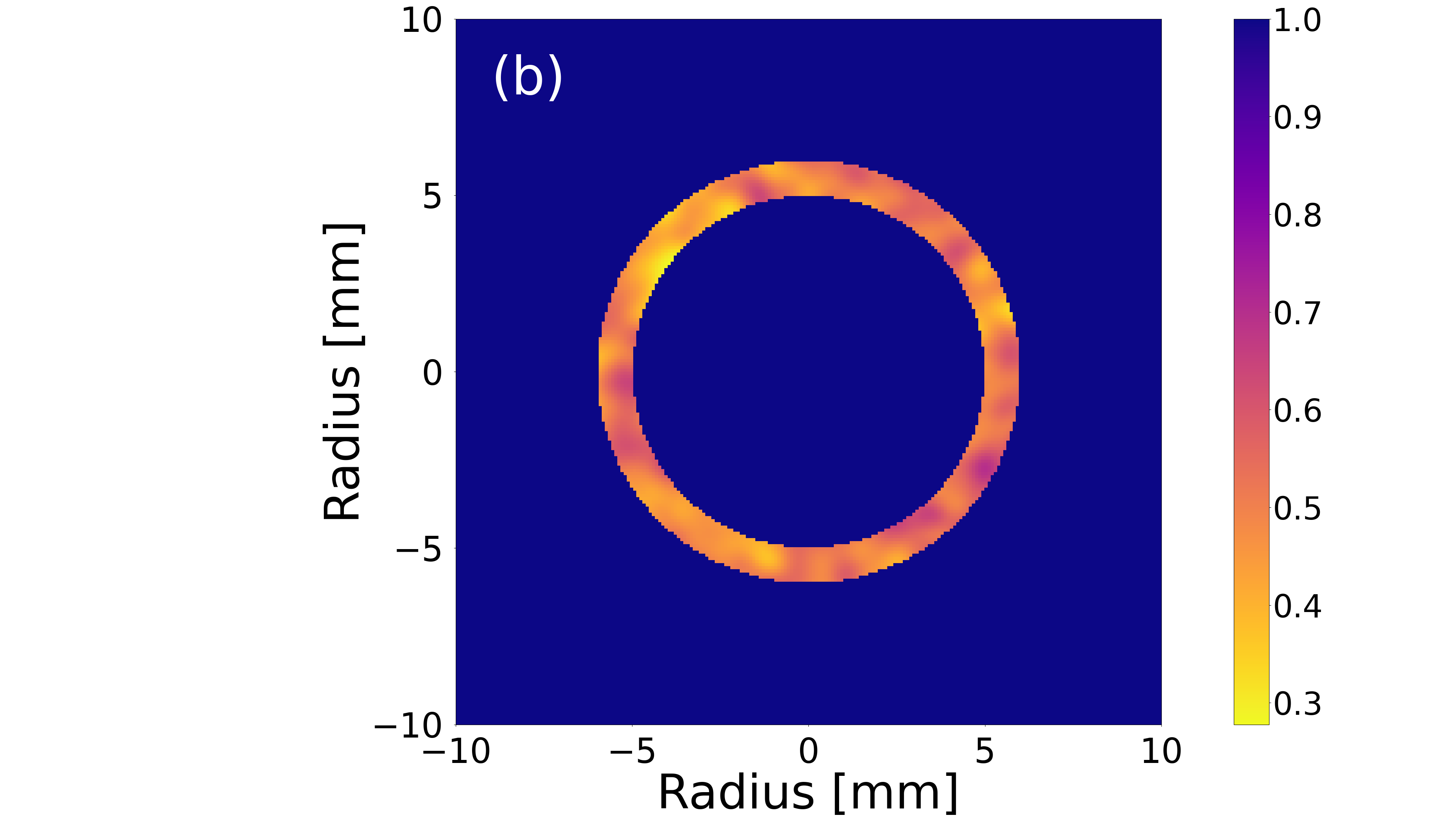}
    \end{minipage}
    \hfill
    \begin{minipage}[t]{0.32\textwidth}
        \centering
        \includegraphics[height=1.2in]{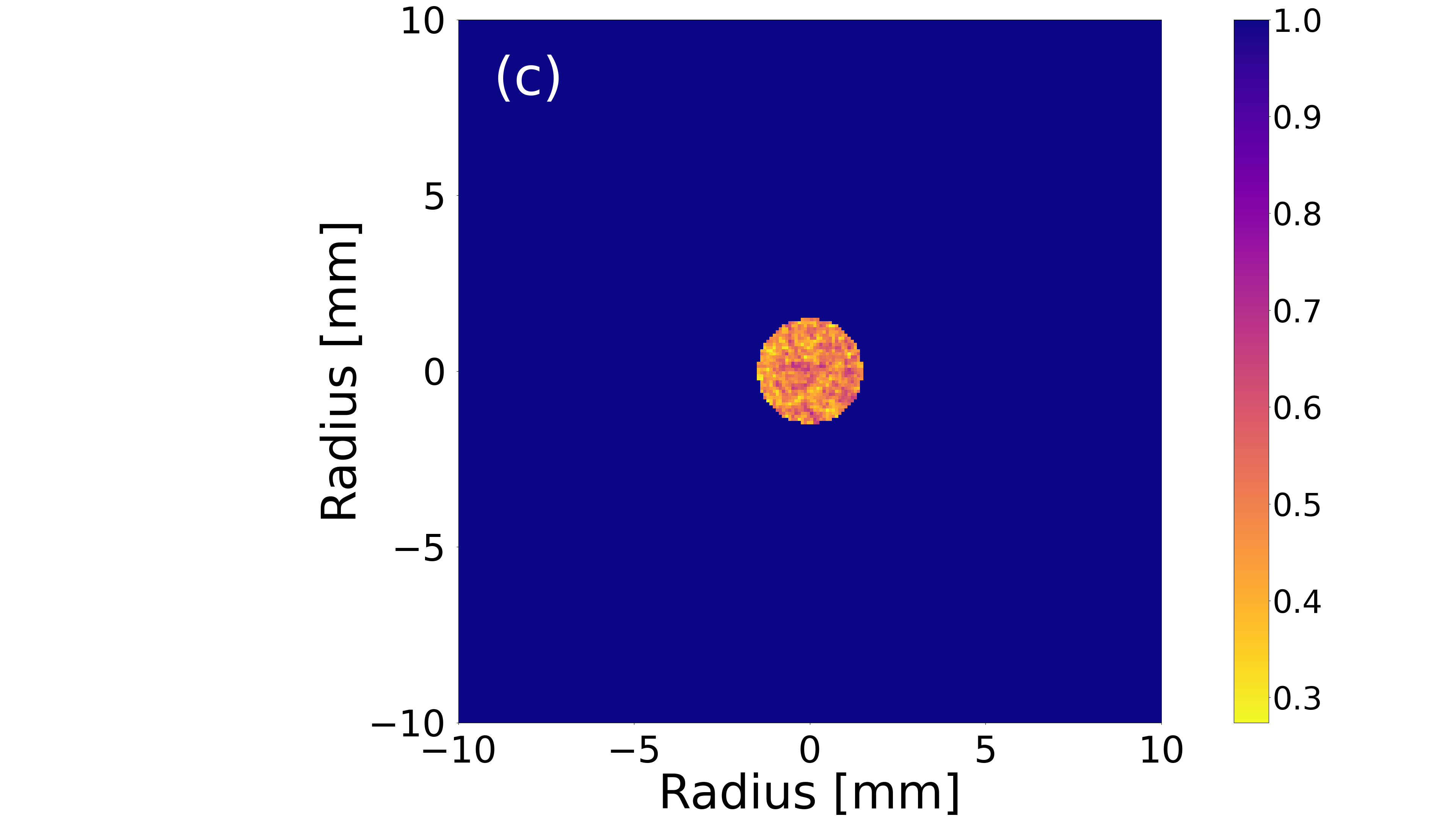}
    \end{minipage}
    \caption{(a, b, c) - the cross-section of the artificially generated turbulent refractive index profile at half-height of the column. The average spatial scales used are (a) $5~\text{mm}$, (b) $1~\text{mm}$, and (c) $100~\text{$\mu$m}$.}
    \label{fig:bpm_density}
\end{figure*}
\par To provide additional insight, we numerically solve the inverse problem, i.e., given the density distribution inside the plasma column, what would be the expected imaging refractometer data? To address this question, we developed the 3D BPM simulation with transparent boundary conditions employing the alternating-direction implicit finite difference method \cite{press1986numerical, okamoto2006}. In this simulation, the 3D refractive index matrix, calculated from the assumed density distribution via plasma dispersion relation for the high-frequency electromagnetic waves, is used as an input (see Fig.~\ref{fig:bpm_density}(a, b, c)), and the complex slow-varying (envelope) electric field is the simulation output. The simulation volume is a 3D box of $40~\text{mm}\times40~\text{mm}\times25~\text{mm}$ with a cylindrical shell at its center. The cylindrical shell diameter approximately matches the plasma column sizes shown in Fig.~\ref{fig:imref_shdw}(b, d, f): $17~\text{mm}$, $10~\text{mm}$, and $3~\text{mm}$ respectively. Cylindrical shell width is about $1~\text{mm}$, which reflects the latest measurements  of the plasma sheath width conducted using Zeeman polarization spectroscopy\cite{Angel2024}. Within this shell, we impose different turbulent density distributions generated separately via an algorithm from Ref.~\cite{timmer_pwrlaw}. We introduce turbulent density distributions with varying average spatial scales to study the density fluctuation scale-related effects. To summarize, the procedure includes generating a turbulent density profile, which is then converted into a refractive index matrix, shown in Fig. 2(a, b, c), for further use as input for BPM simulation runs. Afterward, the BPM simulation is used to propagate a $10~\text{mm}$ single TEM\textsubscript{00} mode laser beam through this cylindrical shell. To generate the synthetic imaging refractometer data shown in Fig.~\ref{fig:bpm_output}(d,e,f), the simulated electric field is Fourier transformed along the vertical direction and its magnitude is taken. The primary objective of these simulation runs is to investigate a trend similar to that observed in Fig. 1(a, c, e), where spatial coherence decreases as the pinch column size decreases, and to determine whether this behavior is consistent with the decreasing fluctuation scale length within the region of interest. Additionally, the code runs aim to explore the possibility of observing vertical widening of the synthetic spectra when using non-turbulent density distributions.

\par The analysis of the imaging refractometer measurements will be in the context of Fourier optics \cite{goodman}, with the primary light-matter interaction being elastic scattering off of electrons in the plasma. We think of an incoming single TEM\textsubscript{00} mode laser beam as a superposition of $N$ plane waves. Before reaching the plasma region, these waves propagate in Fourier space within some angle $\delta \theta$, representing the beam divergence and estimated $\approx\!0.5~\text{mrad}$ for the current research. This narrow $\delta \theta$ spread is the response function of the imaging refractometer when there is no plasma present at the object plane and amounts to a straight line in horizontal direction, which will be referenced later in the text as a $k=0$ line. The response function (preshot) was measured prior to every shot and included in the figure showing the probability distribution of intensities in the laser beam. Relative to this response function, deflection angles will be acquired by each of the individual plane waves propagating through the plasma column and will depend on the electron density gradient along their optical paths via Eq.~\ref{def_angle}.
\begin{eqnarray}
    \theta = \frac{1}{2} \int \frac{\nabla n_e}{n_{cr}} dl \label{def_angle}
\end{eqnarray}
where $\theta$ is the deflection angle, $\nabla n_e$ is the electron density gradient, $n_{cr}$ is the critical plasma density, and $dl$ is the optical path \cite{hutchinson}. The acquired deflection angle is seen as a vertical spread in the imaging refractometer data that indicates local density variations.

\par Fig.~\ref{fig:imref_shdw} presents raw imaging refractometer data at different implosion stages and corresponding shadowgraphy images with the timings relative to the x-ray emission maximum measured by a PCD. The vertical axes in Fig. 1(a, c, e) are labeled with wavenumbers obtained via a calibration procedure outlined in Ref. \cite{RSIpaper}. Note that the mentioned calibration is related to the Fourier plane generated by the first lens of the imaging refractometer setup. Fig.~\ref{fig:imref_shdw}(a) shows an instance when the laser beam size ($\approx9~\text{mm}$) is smaller than the plasma column ($\approx 17~\text{mm}$). The laser field remains almost spatially coherent across the pinch column, indicating the beam interaction with rather smooth plasma electron distribution. Moreover, the spread along the vertical direction of the image is small, suggesting a slight variation in the electron density distribution within the plasma sheath. Most plane waves likely have almost the same optical path through the plasma, keeping the observed lines coherent across the column. In Fig.~\ref{fig:imref_shdw}(b), we show the shadowgraphy image taken almost simultaneously with Fig.~\ref{fig:imref_shdw}(a), which shows a nearly transparent plasma column with developing MRT instabilities at the outside edge.

\par In Fig.~\ref{fig:imref_shdw}(c), we show another phase of the implosion where the laser beam is larger than the plasma, and the $k=0$ mode is being disturbed across the column. However, constant $k$ lines are still present within the measured area. The effects on the laser profile from the MRT instability bubbles are more pronounced but limited to creating interference patterns almost on the background noise level, as the $k=0$ intensity line (left side of Fig.~\ref{fig:imref_shdw}(c) is still the most intense. The electron density distribution inside the MRT bubbles is insufficient to spread the $k=0$ line and produce a sizeable effect likely due to the insufficient density fluctuations. Therefore, the individual plane waves have distinctively different optical paths. The spatial coherence length, measured from the horizontal length of the observed features, depends on the radial position and its max/min values in the $1~\text{mm}$ to $100~\text{$\mu$m}$ range. In Fig.~\ref{fig:imref_shdw}(d), the shadowgraphy pattern reveals that plasma is still primarily transparent; however, some structures of increased line-integrated density are forming across the pinch column, causing some light to be refracted away. Also, the MRT instability bubbles are transparent for the most part except for the edges, which confirms the observations made from the imaging refractometer data where these bubbles seem not to contribute much to a vertical spread.
\par In Fig.~\ref{fig:imref_shdw}(e), we present an image taken about $1 ns$ before the stagnation stage onset, measured by a PCD detector. The radial positions behind the plasma sheath (left- and rightmost radial positions) in Fig.~\ref{fig:imref_shdw}(e), similar to Fig.~\ref{fig:imref_shdw}(d), are mostly identical in terms of their optical path, which manifests itself in the relatively narrow, bright $k=0$ line. Within the plasma column, roughly marked by the white line, is a region where spatial coherence length has dropped dramatically and averages (max/min) from $\approx 200~\text{$\mu$m}$ to $\approx 20~\text{$\mu$m}$. This observation suggests that the individual plane waves are coherent only on a relatively short spatial scale, and their optical paths are significantly different. Moreover, the intensity profile of the zoom-in region marked by a green box is similar to that of laser speckles. The laser speckle effect can be observed when coherent light propagates through a turbulent medium; e.g., see Ref.~\cite{1970speckle}. From the theoretical perspective, the light propagating through the random medium or reflecting off a rough surface is subject to a random walk behavior that randomizes optical paths of the individual plane waves and generates speckles \cite{Goodman1975}.
\par
\begin{figure}[h]
    \includegraphics[width=0.8\linewidth]{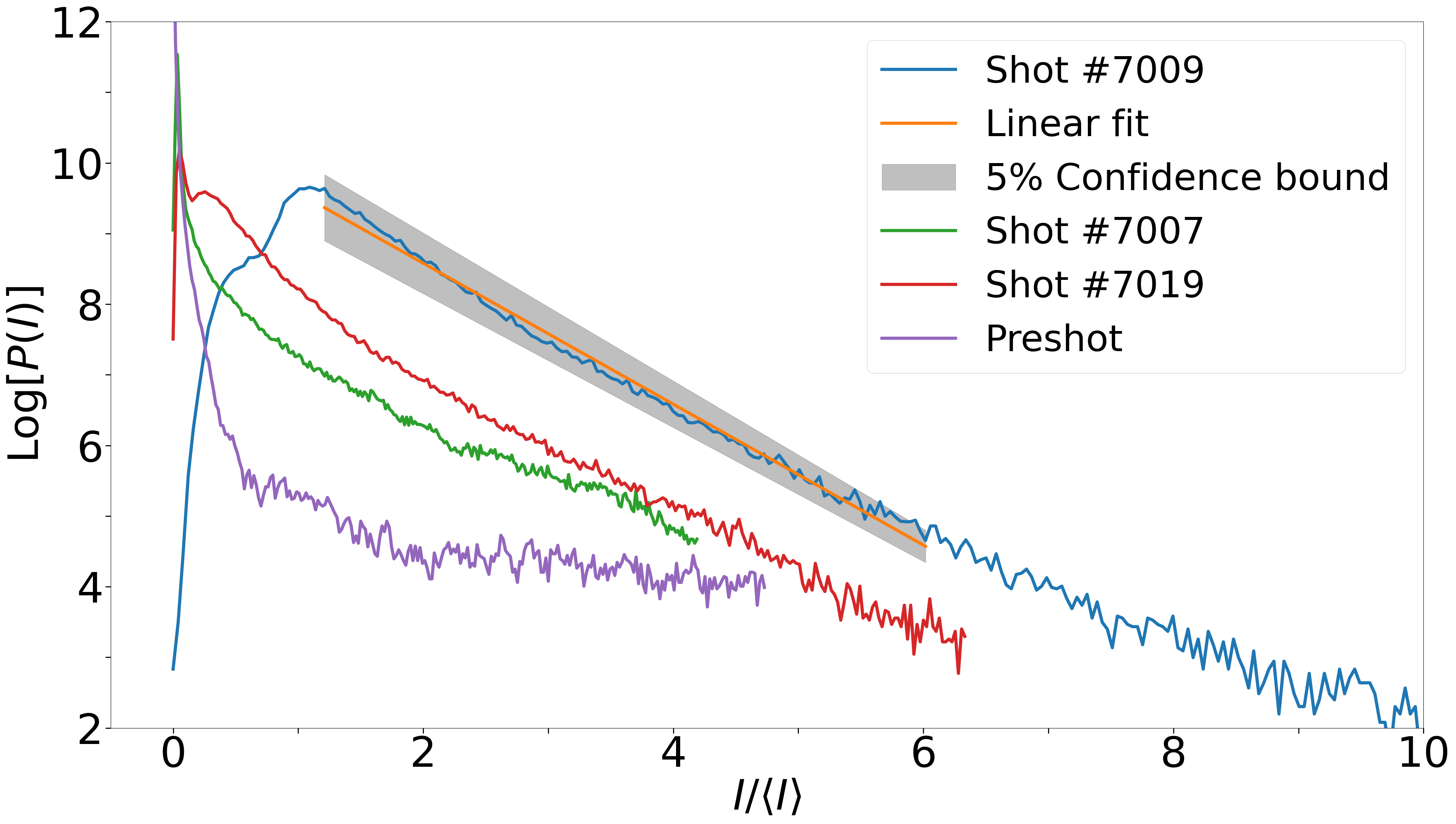}
    \caption{The probability function $P(I)$ on a log scale that the intensity exceeds the value $I$ calculated for the marked regions that appear in Fig.~\ref{fig:imref_shdw}(a,c,e).}
\label{fig:probabD}
\end{figure}

\begin{figure}[h]
    \includegraphics[width=0.8\linewidth]{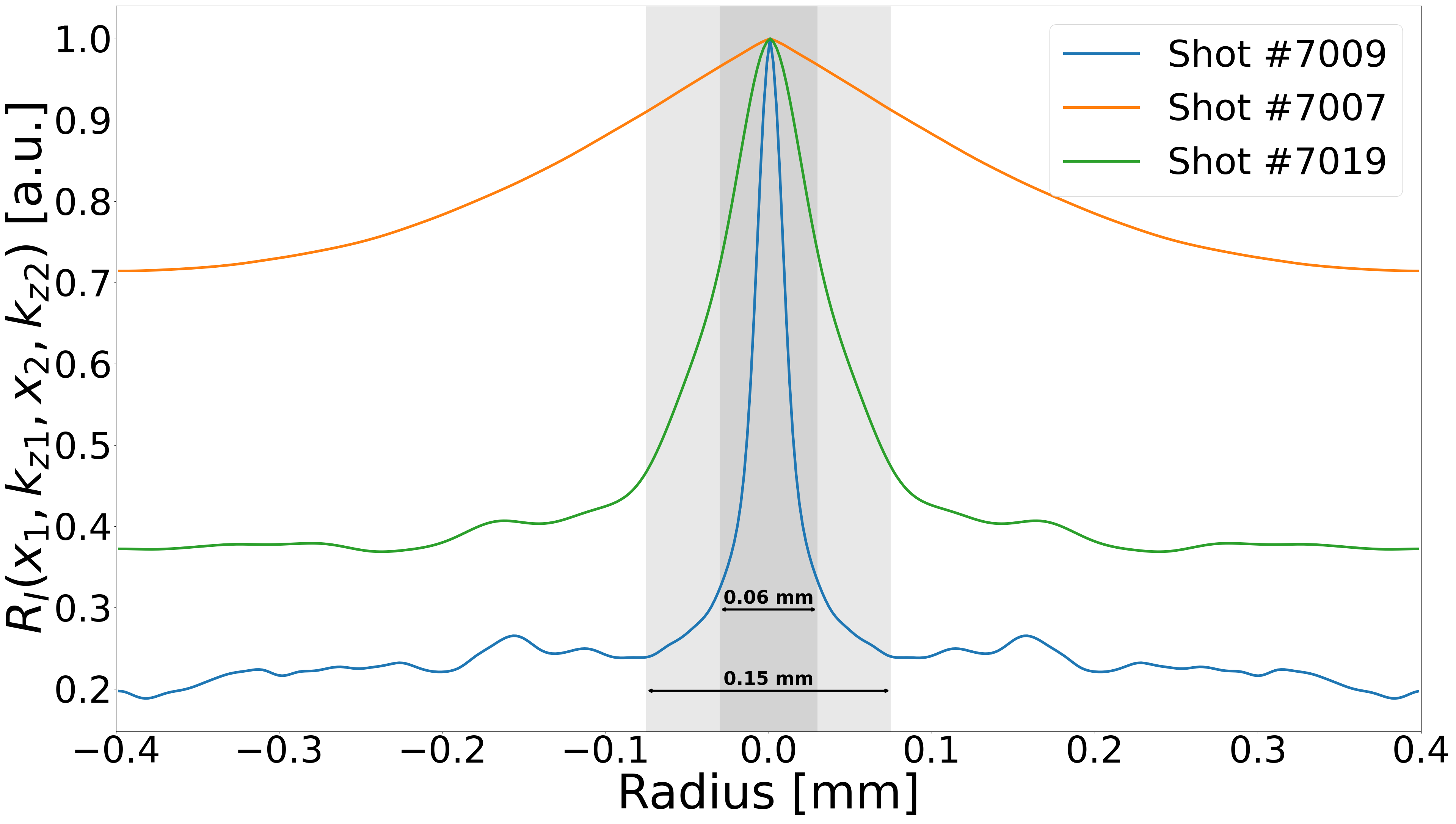}
    \caption{The autocorrelation functions calculated for the speckled regions that appear in Fig.~\ref{fig:imref_shdw}(a,c,e).}
\label{fig:autocor}
\end{figure}

\begin{figure*}[t]
    \centering
    \begin{minipage}[t]{0.32\textwidth}
        \centering
        \includegraphics[width=2in,height=1.2in]{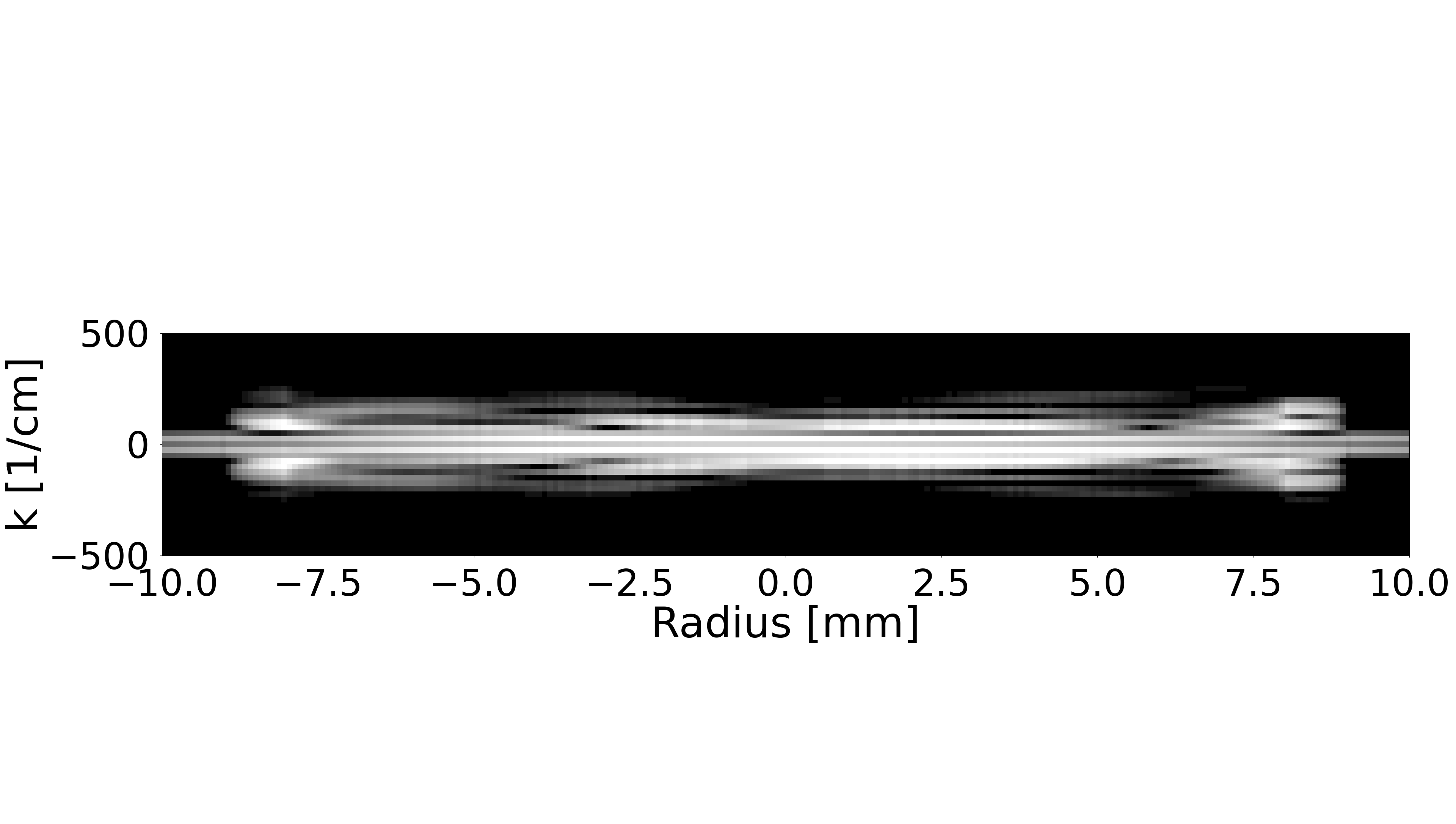}
    \end{minipage}
    \hfill
    \begin{minipage}[t]{0.32\textwidth}
        \centering
        \includegraphics[width=2in,height=1.2in]{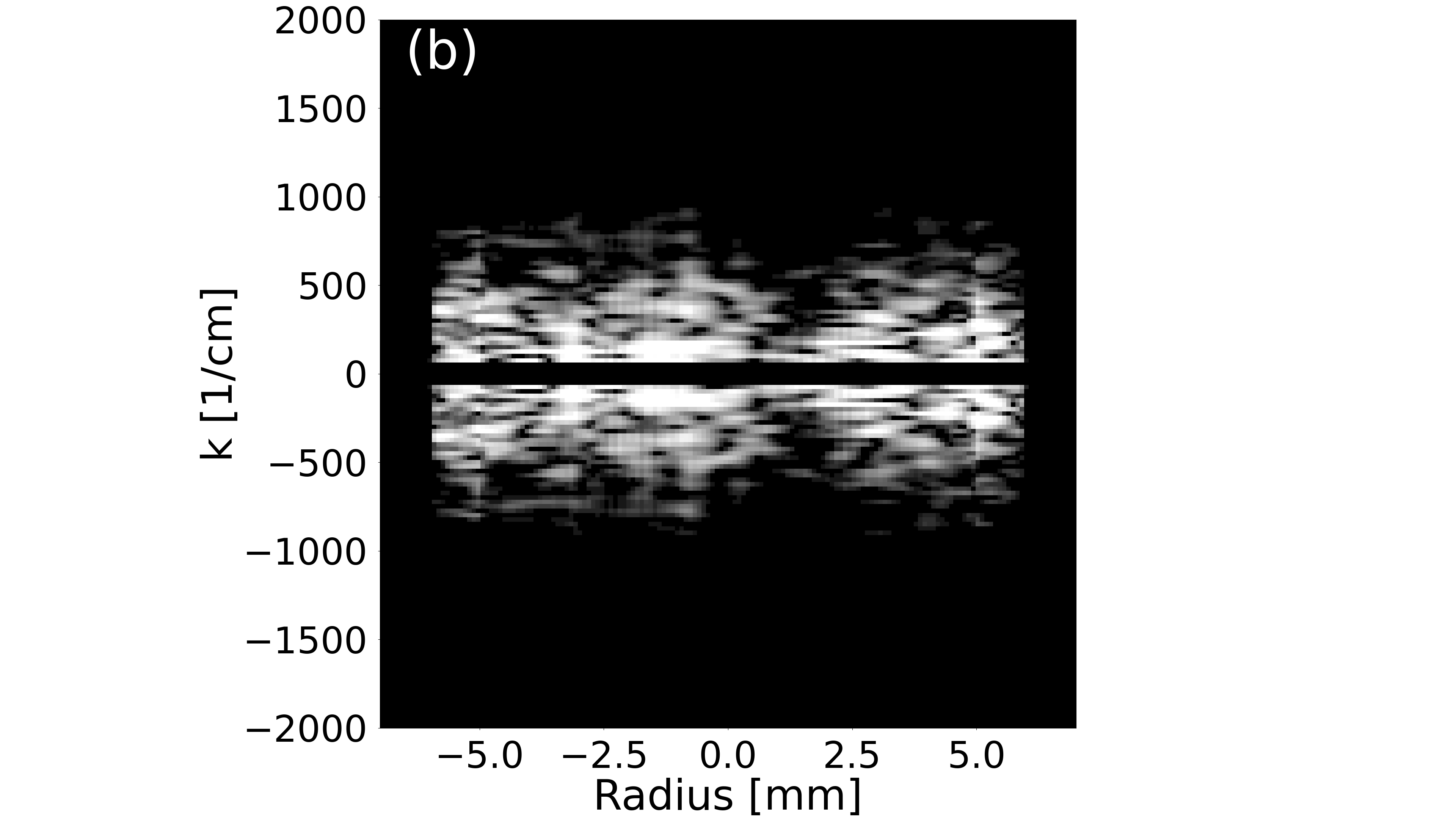}
    \end{minipage}
    \hfill
    \begin{minipage}[t]{0.32\textwidth}
        \centering
        \includegraphics[width=2in,height=1.2in]{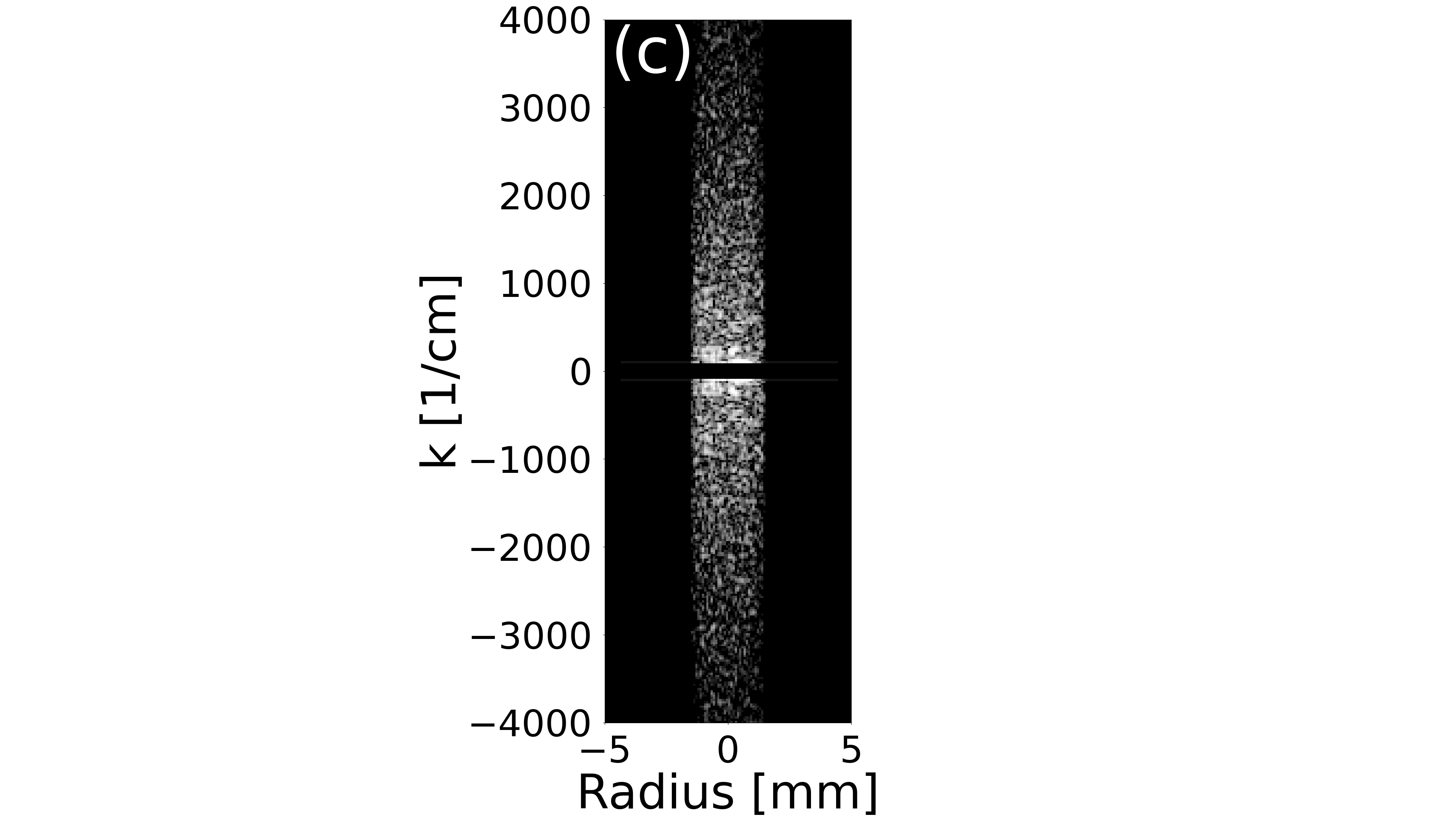}
    \end{minipage}
    \caption{(a, b, c) - The BPM simulated output when the refractive index distributions from Fig. \ref{fig:bpm_density} were used as an input. $k=0$ line has been rescaled and reshaped for clarity.}
    \label{fig:bpm_output}
\end{figure*}

 In Figs. \ref{fig:probabD}, \ref{fig:autocor} we provide more evidence for the individual plane wave phase randomization by computing the intensity distributions' first- and second-order statistics shown in Fig.~\ref{fig:imref_shdw}(a, c, e). Before proceeding, we need to clarify what kind of field we measure. Let $\epsilon (x,y_0,z)$ be a transverse intensity profile of the laser field, which propagates in the direction parallel to the $y$-axis, where $y_0$ is the Fourier plane location of the imaging refractometer optical setup. Since the imaging refractometer utilizes imaging from a 1D optical Fourier transform of the incoming laser field, $\epsilon$, the CMOS sensor is measuring the 1D power spectrum of the laser field per spatial position $x$.
\begin{eqnarray}
    P(I) = \frac{1}{\langle I\rangle} exp{\frac{-I}{\langle I\rangle}} \label{speckles_pdf}
\end{eqnarray}
\par We start by computing the probability function, shown in Eq.~\ref{speckles_pdf}, that the intensity reaches a certain threshold $I$ for the imaging refractometer data presented in Fig.~\ref{fig:imref_shdw}(a, c, e) \cite{Goodman1975}. From the curves seen in Fig.~\ref{fig:probabD} only the blue one (shot \#7009) has a clear linear part and hence, within the confidence bounds, follows the shifted negative exponential distribution. As reported in Ref.~\cite{random_lasers}, the shift may be due to the incoherent background, which is plasma self-emission not entirely suppressed by our optical setup that utilizes $532~nm$ interference filters with $1~nm$ FWHM. Also, the partial depolarization effect of the optical elements on the laser field will lead to the departure from the pure negative exponential distribution \cite{mckechnie2016general}. Yet another factor contributing to the shift is related to the partly blurred image, which in general leads to a different probability function than used here; however, the amount of blurring seems to be small, and speckle contrast relatively high, so this effect is likely to be minor \cite{Goodman1975}. By fitting the linear part of the measured distribution ($\chi^2$ value of $\approx0.9$), the average intensity value of the distribution can be found, which in the polarized speckle should be equal to the standard deviation $\sigma_I$ \cite{Goodman1975}. In the case of Fig.~\ref{fig:imref_shdw}(e), the average intensity computed directly from the data is $\langle I\rangle = 450\pm 50 [a.u.]$ and the standard deviation is $\sigma_I=430 [a.u.]$, which lands the speckle contrast reasonably close to unity. The average intensity value obtained via fitting the linear part of Fig.~\ref{fig:probabD}(blue curve) is $\langle I\rangle=460\pm 40 [a.u.]$ to indicate a good fit. The slight discrepancy in these numbers is due to the abovementioned effects such as background, depolarization, and blurring. The experimental error is mainly due to the CMOS sensor noise level present as a background.

\par Next, we analyze the second-order statistics, the autocorrelation function, which is related to the coherence length of the transverse laser field and indicative of the scale of the density fluctuations present in the plasma column. In Fig.~\ref{fig:autocor}, we show the autocorrelation function, $R_I$, calculated by applying the autocorrelation theorem \cite{goodman}. There is a clear trend of decreasing autocorrelation length, which gradually reduces from being $>0.5~\text{mm}$ for Fig.~\ref{fig:imref_shdw}(a) to as low as $\approx 60~\text{$\mu$m}$ for Fig.~\ref{fig:imref_shdw}(e). The autocorrelation function doesn't decay to zero, which is characteristic of the long-range correlations in the laser field. It seems plausible that the laser field power spectra measurements presented in Fig.~\ref{fig:imref_shdw}(a, c) should contain both short-range correlations driven by the emerging randomness in the optical paths and long-range representing the initially coherent state of the laser field along with mostly coherent scattering within the plasma column. The appearance of weak long-range correlations in the data from shot \#7009 is somewhat unexpected. It is likely related to the analysis region, which includes areas with different properties leading to a non-uniform speckle formation.

\par In Fig.~\ref{fig:bpm_density}(a, b, c), we show the cross-section of different cylindrical targets used to run BPM simulations and in Fig.~\ref{fig:bpm_output}(a, b, c) - the synthetic imaging refractometer data. Different practical density distributions were considered for testing the system's response, including shockwave structures with relatively high density gradients (up to three orders of magnitude jump at the shockwave front) and various plasma sheath widths along the light propagation direction. The investigation reveals that only when sub-mm random transverse density variations are present, there is a sizeable change of the transverse laser field profile in the vertical direction. Also, a similar dynamic observed in the experimental data, i.e., the gradual decrease in spatial coherence length (along the horizontal direction) is clearly seen. In Fig.~\ref{fig:bpm_output}(c), the small-grain fully developed speckle pattern, characterized by the parameters like speckle contrast and standard deviation, can be observed. However, similar statistical analysis is currently unfeasible due to the relatively low resolution of the BPM code where the 2D region of interest has $450\times450$ pixels in it, while the CMOS sensor in our camera has $4210\times6280$ pixels. Nevertheless, analyzing Fig.~\ref{fig:bpm_output} and other images not shown here with $200~\text{$\mu$m}$ average scale, we think it is likely that the average density fluctuation scale in the experiment can be bounded from above at $150~\text{$\mu$m}$ as larger scales produce more radially coherent structures than those observed experimentally. It is understood that an additional research is needed to provide more details. The cylindrically shaped shockwaves, similar to those observed in actual gas-puff z-pinch shots, that produce steep velocity gradients by themselves are incapable of accounting for the experimental results and when simulated are producing only a narrow profile around the $k=0$ line. Therefore, we conclude that to qualitatively reproduce the imaging refractometer data, including laser speckle and the trend of decreasing spatial coherence of a laser field with decreasing pinch column size, the necessary condition is to have sub-mm scale density fluctuations. Also, Fig.~\ref{fig:bpm_output}(a) suggests it is likely that turbulence in gas-puff z-pinches may be present almost from the start of the current due to some level of consistency between the experimental and synthetic data generated with large-scale ($5~mm$) random density fluctuations.
\par To summarize, we have presented imaging refractometer data that shows the laser speckle generation, necessitating random electron density fluctuations. By analyzing the data, we demonstrate that speckles are generated as the laser light propagates through the plasma sheath. By studying the measured intensity distribution, we identify the main features attributed to speckles: the distribution's shape, speckle contrast, and autocorrelation. These conclusions are supported by subsequent BPM simulations, where only random fluctuations in the transverse direction can reproduce the data and show the experimentally observed trend of decreasing radial coherence of the laser field with a decreasing average spatial scale of density fluctuations. The imaging refractometer data provides the first direct experimental evidence of turbulence in gas-puff z-pinch implosions. Future work will focus on the exact turbulent transition mechanism, including studies involving other gas species, additional investigation of the early implosion stages where BPM simulations suggest turbulence, and the potential dependence of synthetic imaging refractometer data on the specific probability density function used to generate density distribution.

\begin{acknowledgments}
The authors would like to thank Todd Blanchard, Daniel Hawkes, and Harry Wilhelm for their support in conducting this research. This research was supported by the Cornell Laboratory of Plasma Studies, by the Engineering Dean's office through the College Research Incentive Program, by the K. Bingham Cady Memorial Fund, and by the Air Force Office of Scientific Research under award number FA9550-24-1-0066.
\end{acknowledgments}

\section*{Data Availability Statement}
The data that support the findings of this study are available from the corresponding author upon reasonable request.

\end{document}